\documentclass[12pt,a4paper]{article}
\usepackage[T1]{fontenc}
\usepackage[utf8]{inputenc}
\usepackage{authblk}
\usepackage{amsmath,amssymb,epsfig,cite,dsfont}
\usepackage{array,multirow}
\usepackage {color}
\usepackage[toc,page]{appendix}
\numberwithin{equation}{section}
\usepackage{afterpage}
\usepackage{cite}
\usepackage[utf8]{inputenc}
\usepackage[english]{babel}
\usepackage{amsmath}
\usepackage{amsfonts}
\usepackage{amssymb}
\usepackage{makeidx}
\usepackage{graphicx}
\usepackage[left=2cm,right=2cm,top=2cm,bottom=2cm]{geometry}
\usepackage{color}

\title{
\vspace{-2.2cm}
\begin{center} \bf{Covariant Jacobi Brackets for test Particles
\linethickness{.05cm}\line(1,0){433}
}\end{center}}
\author[1]{ M. Asorey
}
\author[2,3]{F. M. Ciaglia
}
\author[2,3] {F. Di Cosmo 
}
\author[4,5] {A. Ibort}
\author[2,3]{ G.  Marmo,
}
\affil[1]{\small Departamento de F\'{\i}sica Te\'orica, 
Universidad de Zaragoza, Pedro Cerbuna 12, 50009 Zaragoza, Spain.}
\affil[2]{\small  Dipartimento di Fisica {\sl E. Pancini} dell'Universit\`a  {\sl Federico II} di
Napoli, 80126 Napoli, Italy. }
\affil[3]{\small  INFN Sezione di
Napoli, Complesso Universitario di Monte S. Angelo, via Cintia, 80126 Napoli, Italy.}
\affil[4]{\small Departamento de Matem\'aticas, Universidad Carlos III de Madrid,
Avda. de la Universidad 30, 28911 Legan\'es, Madrid, Spain.}
\affil[5]{\small ICMAT, Instituto de Ciencias Matem\'aticas (CSIC - UAM - UC3M - UCM),
Nicol\'as Cabrera,13–15, Campus de Cantoblanco, UAM, 28049 Madrid, Spain.}
\date{}

\begin{document}

\maketitle

\begin{abstract}
We show that the space of observables of test particles carries a natural Jacobi structure which is manifestly invariant under the action of the Poincar\'{e} group. Poisson algebras may be obtained by imposing further requirements. A generalization of Peierls procedure is used to extend this Jacobi bracket on the space of time-like geodesics on Minkowski space-time.
 
\end{abstract}

\section*{Introduction}
Relativity, Special and General, and Quantum Mechaics are two major conquests of the physics of the previous century. A general theory which would be able to incorporate both of them is still lacking and it is one of the most challenging and fascinating open problems of modern theoretical physics. 

The advent of relativity has also forced physicist to think of dynamics in different terms. In the case of special relativity Dirac \cite{Dirac} explains very well what would be a resonable meaning of relativistic dynamics: the construction of a set of observables closing on the Lie algebra of the Poincar\'{e} group. In order to preserve covariance with respect to this group, Dirac suggests to replace the conventional time translation with the full translation subgroup of the Poincar\'{e} group and various realizations of the Poincar\'{e} algebra are called \textit{forms} of relativistic dynamics. 

In modern mathematical language we would say that a form of relativistic dynamics consists in a realization of the abstract commutation relations of the Lie algebra of the Poincar\'{e} group as derivations of a certain associative algebra of observables. Notice that this formulation does not discriminate between classical and quantum systems and can be used in both instances. Indeed one of the most relevant achievement of modern mathematics as applied to Physics has been the possibility of using abstract relations to understand and discuss physical laws without using a specific realization. This way of formulate problems allows to separate the aspects which are general from the ones that are related with a chosen realization. For instance the language of abstract algebras is at the base of Dirac's \textit{analogy principle} that allows to relate the quantum description of physical systems and their corresponding classical counterparts. 

In a two pages paper Wigner raised the question\cite{Wigner}: to what extent the equations of motion determine the commutation relations? 
Adopting Dirac's point of view ,one might rephrase Wigner's problem by searching to what extent the invariance under the Poincar\'{e} group would determine the commutation relations.

As the Lie algebra structure is essentially determined  by the associative structure which admits the Lie algebra as an algebra of derivations \cite{GM},this amounts to ask for all possible associative products which admit as derivations the infinitesimal generators of the Poincar\'{e} group. Obviously the answer is not unique: for instance, in the quantum setting the two associative operator products $A \cdot B$ and $A\cdot K \cdot B$ where $K$ is any operator function of the Casimir operator for the Poincar\'{e} goup, admit the Lie algebra of the Poincar\'{e} group as derivations. Nevertheless in both cases these derivations are inner (see Ref. \cite{GM}), i.e. they are realized as commutators with an element of the algebra itself. Actually, in the case of the algebra of observables of a quantum mechanical system, all the derivations are inner. However the classical limits (for instance in the sense of Moyal products see Ref. \cite{EMM}) of these two associative products and of the corresponding commutators have different properties; in the first case the commutator is related to a Poisson bracket, whereas for the second situation the limiting algebra defines a Jacobi bracket. \footnote{More general in paper \cite{EMM} it is shown that in presence of KMS states, the classical limit is indeed a Jacobi bracket.}
Even if these two structures have different properties (as we will see in the following sections) they both emerge as classical limits of two alternative solutions to the same algebraic problem. Since we believe that Quantum Mechanics is a fundamental theory and the corresponding classical description should describe \textit{emergent} structures we cannot exclude the possibility of using Jacobi brackets to describe the algebra of observables of a classical system.     

In this letter we would like to investigate this possibility: in particular we will show that in some situations of physical interest the space of observables of a theory carries a natural Jacobi structure, whereas Poisson algebras can be obtained by imposing further requirements. Moreover it will be shown that such Jacobi bracket is manifestly covariant under the action of the Poincar\'{e} group and is the most general local Lie algebra structure (according to Kirillov theorem (see Ref. \cite{Kirillov})) possessing this property.
More specifically, the family of examples related with the description of the motion of
relativistic test particles with mass $m$ will be used to illustrate previous ideas. Then it will be shown how the description of covariant brackets started by Peierls \cite{Peierls} and consistently developed in a geometrical context in a recent contribution \cite{Asorey} becomes a particular instance of the present construction. In this respect, the present letter should be considered to be an updating of previous one. We restrict here to test particles to avoid unnecessary technicalities. 

\section{Contact Manifolds out of the Klein-Gordon equation}
In this section we show how a Jacobi bracket emerges naturally when considering Klein-Gordon equation.
In a given standard linear coordinate system $(x_0,\, x_1,\, x_2,\, x_3)$ for Minkowski space-time, Klein-Gordon equation is usually written as
\begin{equation}
\left( \dfrac{\partial ^2}{\partial x_0^2} - \nabla^2 + \dfrac{m_0^2c^2}{\hbar ^2} \right) \psi = 0 , 
\end{equation}
where $\nabla^2 = \frac{\partial ^2}{\partial x_1^2} + \frac{\partial ^2}{\partial x_2^2} + \frac{\partial ^2}{\partial x_3^2}$ is the Laplacian operator.

Often, Klein-Gordon equation is written in the following compact form
\begin{equation}
\left( \square + m^2 \right)\psi = 0 \,,
\end{equation}
where $\square=\dfrac{\partial ^2}{\partial x_0^2} - \nabla^2$ is the d'Alembert operator.

This equation was proposed by Oskar Klein and Walter Gordon to describe relativistic electrons \cite{Lizzi}. However, as electrons carry also a spin, the equation does not provide with a satisfactory description of them and it is suitable only for spinless particles. For instance it could describe composite particles like pions, or Higgs bosons.

It is possible to write the equation without using coordinates by means of the exterior differential and the codifferential, $d$ and $\delta$ respectively, say 
$$
\delta d \psi = m ^2 \psi\,.
$$
This form shows very clearly that no time and space splitting of space-time is required to formulate Klein-Gordon equation, see Ref. \cite{CDLM}

By associating a ``symbol'' to this differential operator by means of the functions $e^{\pm i p_{\mu}x^{\mu}}$, see Ref. \cite{EMS}, we would find a dispersion relation for the momentum four-vector of the relativistic particle 
\begin{equation}
e^{\pm i p_{\mu}x^{\mu}} \square e^{\mp i p_{\mu}x^{\mu}} = p_{\mu}p^{\mu}\,.
\end{equation} 
This dispersion relation determines a seven-dimensional submanifold $\Sigma_m$ of the phase space $T^*\mathbb{R}^4$ as follows
\begin{equation}
\Sigma_{m} \equiv \left\lbrace \left( x,\,p \right) \vdash p^{\mu}p_{\mu} = m^2 \right\rbrace\,.
\end{equation}

Denoting with $i_{\Sigma}$ the canonical immersion of $\Sigma_m$ into $T^*\mathbb{R}^4$, we have that the pull-back $i_{\Sigma}^*(\theta_0) = \theta_{m}$ of the natural Liouville one-form $\theta_0 = p_{\mu}dx^{\mu}$, defines a contact structure (see section 3). 
It is worth noticing that both $\theta_0$ and the submanifold $p^{\mu}p_{\mu} = m^2$ are manifestly Poincar\'{e} invariant.
Having obtained a contact structure out of the Klein-Gordon equation, we will show how to get a Lie bracket associated with it.

\section{Contact Manifolds and Jacobi Brackets}
Contact manifolds are usually presented as the odd-dimensional counterpart of sympectic manifolds. If we look from the dual point of view of functions\cite{CDLM}, the space of functions on a symplectic manifold can be equipped with a canonical Poisson bracket whereas the space of functions on a contact manifold can be endowed with a ``natural'' Jacobi bracket. The unique role of the Jacobi bracket originates from a theorem by Kirillov which examines the most general Lie bracket one may define on the algebra  of functions when an additional locality reuirement is assumed,locality meaning that $supp\left( \left[ f,g \right] \right) \subseteq supp(f) \cap supp(g)$.  
Let us begin with some definitions. 
   
On some $(2n+1)$-dimensional manifold $\mathcal{M}$, a differential form $\theta$ defines a contact strucure if $\theta \wedge (d\theta)^n \neq 0$, i.e., it is a volume element. However if we multiply such a form by a never vanishing function we get another one form satisfying the same condition. Therefore a contact structure is actually defined as an equivalence class of one forms related by multiplication by a never vanishing function. By means of this contact one form we are selecting hyperplanes of the tangent space to the manifold which are called contact elements (for more details see Ref. \cite{Arnold}). 

This arbitrariness may be reduced by imposing invariance requirements, in our case it would be invariance with respect to the Poincar\'{e} group.

A manifold $\mathcal{M}$ endowed with a contact structure, is called a contact manifold. 
Given a contact manifold we define a Lie algebra structure on the space of functions by means of the following formula
\begin{equation}
\left[ f,\, g \right]\theta \wedge (d\theta)^n = (n-1)df \wedge dg \wedge \theta \wedge(d\theta)^{n-1} + (f dg - g df)\wedge (d \theta)^n \,. 
\label{Jacobi bracket}
\end{equation}   
This bracket is clearly local by construction, and satisfies the Jacobi identity
$$
\left[ \, f \, , \, \left[ \, g \, , \, h \, \right] \, \right] = \left[ \, \left[ \, f\, , g \, \right]  , \, h \, \right] + \left[ \, g \, , \, \left[ \, f \, , \, h \, \right] \,  \right]
$$
which expresses the property of defining a derivation of the Lie product. It should be stressed that the bracket only depends on the one form.Therefore it will possess all the invariance properties enjoyed by $\theta$.

To make contact with the usual definition of Jacobi bracket we define a vector field $\Gamma$ (also called Reeb vector field) and a bivector field $\Lambda$ with the help of $\theta$ and $d\theta$, satisfying the following properties
\begin{equation}
i_{\Gamma} \theta \wedge (d\theta)^n = (d\theta)^n 
\label{i_G T0}
\end{equation}
\begin{equation}
i_{\Lambda} \theta \wedge (d\theta)^n = n \theta \wedge (d\theta)^{n-1}\, .
\label{i_L T0}
\end{equation}
Previous bracket may be given now in the more conventional form by setting
\begin{equation}
\left[ f \, , \, g \right] = \Lambda(df , dg) + fL_{\Gamma}g - gL_{\Gamma}f \,, 
\label{Jacobi bivector}
\end{equation}
where $L_{\Gamma}$ stays for the Lie derivative along $\Gamma$.
Jacobi identity in this case corresponds to the following requirements on the couple $\left( \Lambda, \Gamma \right)$:
\begin{equation}
\left[ \Lambda, \Lambda \right]_S = 2\Gamma \wedge \Lambda
\label{L,L}
\end{equation}
\begin{equation}
L_{\Gamma}\Lambda = 0 
\label{L_G L}
\end{equation} 
where the bracket $\left[\cdot, \cdot \right]_S$ is the Schouten brackets on the algebra of multivctors on a manifold (see Ref. \cite{GM2}). 

With any function $f$ we can associate a first-order differential operator 
$$
\tilde{X}_f = \Lambda(df, \cdot) +f\Gamma - L_{\Gamma}f \,,
$$
and it is worth pointing out that the identity function is not mapped onto $0$ but gives the vector field $\Gamma$.

Notice that Leibniz rule is replaced by 
$$
\left[ f , gh \right] = \left[ f, g \right] h + g \left[ f,h\right] - \left[ f, 1 \right] gh
$$
which explains the difference between Jacobi brackets and Poisson brackets. This generalized Leibniz rule says that the bracket is actually associated with a bidifferential operator instead of a bivector field like in the case of the Poisson brackets. 

In general we define the Hamiltonian vector field, $X_f$, associated with the function $f$, to be the vector field 
\begin{equation}
X_f = \Lambda(df , \cdot ) + f\Gamma \,,
\end{equation}
and this association is a homomorphism of Lie algebra, i.e. 
$$
\left[ X_f , X_g \right] = X_{\left[ f,g \right]}\,.
$$
We deduce immediately that on the subalgebra of functions $f$ such that $L_{\Gamma}f=0$, the Jacobi bracket becomes a Poisson Bracket.

Before ending this section let us notice that the definition $\eqref{Jacobi bivector}$ of a Jacobi structure by means of a bivector field $\Lambda$ and a vector field $\Gamma$ satisfying properties $\eqref{L,L}$ and $\eqref{L_G L}$ is independent from the existence of a underlying contact manifold. It is worth noticing that this definition shows that a Jacobi bracket is unrelated to the dimensions of the manifold. Furthermore this definition fits better in the case of infinite dimensional manifolds where volumes are harder to define.

\section{Jacobi brackets associated with Klein-Gordon equation}
Let us come back to the submanifold $\Sigma_{m} \subset T^*\mathbb{R}^4$ presented in Section 1. As already said, the pull-back of the one form $\theta_0 = p_{\mu}dx^{\mu}$ to this submanifold defines a contact structure, say $\theta_m$, as can be seen by direct computations.
Eventually we get:
\begin{eqnarray*}
&\theta_m\wedge (d\theta_m)^3 = (p^3 dp^0 \wedge dp^1 \wedge dp^2 - p^2 dp^0 \wedge dp^1 \wedge dp^3 + p^1 dp^0 \wedge dp^2 \wedge dp^3 + \\
&+ p^0 dp^1 \wedge dp^2 \wedge dp^3)\wedge dx^0 \wedge dx^1 \wedge dx^2 \wedge dx^3 \neq 0 \,.
\end{eqnarray*}  

As explained in the previous section, the contact structure is defined up to multiplication by a non vanishing function. Using as a reference one form the canonical Liouville one form, we can consider all the family we may obtain by means of a conformal factor. The adapted lifting of the Poincar\'{e} algebra from the configuration space to the phase space is obtained by requiring that the associated vector fields on the phase space preserve the chosen one form. The Lie algebra on which the lifted vector fields will close is the Poincar\'{e} algebra itself if the conformal factor is function of a Casimir. In particular, if the conformal factor is function of the Casimir $p^2 = p_{\mu}p^{\mu}$, then, on the submanifold $\Sigma_m$ it becomes actually a constant. This means that the requirement of Poincar\'{e} invariance for the potential one-form on $T^{*}\mathbb{R}^4$ selects a particular contact form $\theta_m$ on $\Sigma_m$ which is unique apart from a multiplicative constant. 

Let us now introduce the Jacobi bracket associated with the above contact form. As we can see from the definition $\eqref{Jacobi bracket}$ this bracket is entirely defined in terms of the exterior differential $d$ and the contact one-form $\theta_m$, hence, being these ingredients invariant with respect to the Poincar\'{e} group, the Jacobi bracket itself will be fully invariant with respect to Poincar\'{e} group. 

As explained in section 1, we may write the Jacobi bracket in terms of a suitable pair $(\Lambda_{\Sigma},\,\Gamma )$. At this purpose, let us consider the following tensor fields on $T^{*}\mathbb{R}^4$
\begin{equation}
\Lambda_{\Sigma} =\left( g^{\mu \nu} - \dfrac{p^{\mu}p^{\nu}}{m^2} \right) \dfrac{\partial}{\partial p^{\mu}}\wedge \dfrac{\partial}{\partial x^{\nu}}
\label{L_S}
\end{equation} 
and 
\begin{equation}
\Gamma = \dfrac{p^{\mu}}{m^2} \dfrac{\partial}{\partial x^{\mu}}\, .
\end{equation}

It is easy to verify that even if we have used a set of coordinate functions of the full phase space, say $(x^{\mu},p^{\mu})$, these two tensors are actually written in terms of vector fields which are tangent to the submanifold $\Sigma_m$ and therefore belong to the tensor fields built out of the tangent bundle of $\Sigma_m$. This follows because both of them vanish when contracted with the differential of the Casimir function.
 
Direct computations show that these two tensors satisfy all the properties discussed in the previous section. Therefore we can introduce the following bracket on the set of differentiable functions on $\Sigma_{m}$:
\begin{equation}
\left[ f , g \right] = \Lambda_{\Sigma}(df, dg) + fL_{\Gamma}g - gL_{\Gamma}f\,.
\end{equation}
According to this bracket, for instance, the four space and time functions do not commute. At the quantum level they would not describe localization on space-time. Indeed we get the following commutation relations
\begin{eqnarray}
&\left[ x^{\rho}, x^{\sigma} \right] = \dfrac{x^{\rho}p^{\sigma}-x^{\sigma}p^{\rho}}{m^2} \\
&\left[ p^{\rho} , x^{\sigma}\right] = g^{\rho \sigma} \\
&\left[ p^{\rho} , p^{\sigma}\right] = 0 \, .
\label{Commutation Relations}
\end{eqnarray}

Let us point out once more that both tensors fields $\lambda_{\Sigma}$ and $\Gamma$ are invariant under the Poincar\'{e} group $\mathcal{P}$. Therefore the associated Lie algebra $\mathfrak{p}$ acts as an algebra of derivations for the Jacobi bracket and maps the subalgebra of functions which are invariant under $\Gamma$ into itself. Because $\mathfrak{p}$ also preserves $\Lambda_{\Sigma}$ it is also an algebra of derivations for the Poisson subalgebra associated with $\Lambda_{\Sigma}$ and, consequently, it may be realized in terms of Hamiltonian vector fields associated with the conventional generators $M_{\mu}^{\nu} = x_{\mu}p^{\nu}-x_{\nu}p^{\mu}$ and $p^{\mu}$.

The corresponding Hamiltonian vector fields in the sense of the Jacobi structure are the vector fields 
\begin{eqnarray}
&X_{\rho \sigma} = q_{\rho}\dfrac{\partial}{\partial x^{\sigma}}- q_{\sigma}\dfrac{\partial}{\partial x^{\rho}} + p_{\rho}\dfrac{\partial}{\partial p^{\sigma}} - p_{\sigma}\dfrac{\partial}{\partial p^{\rho}} \\
&X_{\mu} = \dfrac{\partial}{\partial x^{\mu}} \,, 
\end{eqnarray} 
and they coincide with the evaluation on the submanifold $\Sigma_m$ of the generators of the canonical action of the Poincar\'{e} group on $T^*\mathbb{R}^4$ with respect to the symplectic structure $\omega = d\theta_0$. 

It would be possible to realize this algebra also in terms of Hermitian operators, acting on square integrable functions on space-time. We shall not enter, however, in the physical intepretation of this realization to avoid facing all problems connected with the definition of time-operator, see Ref. \cite{Ciaglia}.

A final remark before moving to the discussion of two-point commutation relations. Our construction is quite general and could be dealt with in general abstract terms and for scalar operators not restricted to be second order. It would also be possible to consider Dirac-like operators by using the formulation in terms of Dirac-Kahler differential operators on differential forms (for more details see Ref. \cite{Zampini}). However this would take us too far way from the main stream of our letter.

\section{Klein-Gordon and two-point Jacobi brackets}
Here we will define a two-point Jacobi bracket starting from the Klein-Gordon equation.

Given a differential operator of order $k$ it is possible to extract from it a PDE of Hamilton-Jacobi type \cite{MMM}. Indeed let us consider a $k$-order differential operator $D^{(k)}$ on some configuration manifold $Q$. Since $D^{(k)}$ is of order $k$, the $k^{th}$-commutator 
\begin{equation}
\left[ \cdots , \left[D^{(k)} , \hat{S} \right],\cdots, \hat{S} \right]
\label{H-J symbol}
\end{equation}
is a multiplication operator. If we call $f_D$ the polynomial functions on $T^*Q$ associated with a differential operator as principal symbol, the resulting multiplication operator in $\eqref{H-J symbol}$ coincides with the pull-back of $f_D$ through the differential $dS$. Therefore we can write a Hamilton-Jacobi type equation according to the following formula:
\begin{equation}
\left[ \cdots , \left[D^{(k)} , \hat{S} \right],\cdots, \hat{S} \right] = (dS)^*(f_D) = c \,.
\end{equation}
For the Klein-Gordon differential operator we get the equation
\begin{equation}
g^{\mu \nu}\dfrac{\partial S}{\partial x^{\mu}} \dfrac{\partial S}{\partial x^{\nu}} = m^2\,,
\end{equation}
where we have chosen the value $m^2$ for the constant $c$. We can notice that this coincide with the Hamilton-Jacobi equation associated with the Hamiltonian function $H=p_{\mu}p^{\mu}$ on $T^*\mathbb{R}^4$.  

A complete solution of this equation is provided by the function 
\begin{equation}
S(x,y)= k_{\mu}(x - y)^{\mu}
\label{H_J Solution}
\end{equation}
with $k_{\mu}$ parameters which satisfy the requirement $k_{\mu}k^{\mu} = m^2$. 
For instance, we could take the components $p_{\mu}$ of the four-momentum as parameters. By means of the evolution map $\phi_{\tau}$ we can associate to each point of the cotangent bundle $\xi= (x^{\mu}, p_{\mu})$ the pair of points of $\mathbb{R}^4$ $\eta = (x^{\mu}, y^{\mu})$, such that 
\begin{equation}
y^{\mu} = \phi_{\tau=1}^{\mu}(\xi)\,.
\end{equation} 
In this expression $\phi_{\tau}^{\mu}$ is the $\mu$-th space-time coordinate of the flow associated with the evolution generated by $H$ corresponding to the initial condition $\xi$. In particular we select the point which corresponds to the value $\tau = 1$ of the parameter of the solution. For this simple situation we get 
\begin{equation}
p_{\mu} = g_{\mu \nu}(y^{\nu} - x^{\nu})\,,
\end{equation} 
where $g_{\mu \nu}$ is Minkowski metric.

If we replace this expression in the solution $\eqref{H_J Solution}$ we obtain the following two-point function 
\begin{equation}
S(x,y) = (x-y)_{\mu}(x-y)^{\mu}\,.
\end{equation}
The submanifold $\tilde{\Sigma}_m$ of $\mathbb{R}^4\times \mathbb{R}^4$ satisfying the dispersion relation $g^{\mu \nu}\dfrac{\partial S}{\partial x^{\mu}} \dfrac{\partial S}{\partial x^{\nu}} = m^2$ is an odd dimensional submanifold where the one-form $\tilde{\theta} = dS^*(\theta_0)$ defines a contact structure. Computations which are analogous to the ones in the previous section allow to define the following Jacobi bracket on the submanifold $\tilde{\Sigma}_m$ by means of the two following tensor fields:
\begin{eqnarray*}
& \tilde{\Lambda} = \left( g^{\mu \nu}- \dfrac{(x-y)^{\mu}(x-y)^{\nu}}{m^2} \right)\dfrac{\partial }{\partial x^{\mu}}\wedge \dfrac{\partial}{\partial y^{\nu}} \\
& \tilde{\Gamma} = \dfrac{(x-y)^{\mu}}{m^2}\left ( \dfrac{\partial}{\partial x^{\mu}} + \dfrac{\partial}{\partial y^{\mu}} \right) \,.
\end{eqnarray*}
The associated commutation relations are:
\begin{eqnarray*}
&\left[ (x+y)^{\rho}, (x+y)^{\sigma}  \right] = \dfrac{2}{m^2}\left( (x+y)^{\rho}(x-y)^{\sigma}- (x+y)^{\sigma}(x-y)^{\rho} \right)\\
& \left[ (x+y)^{\rho}, (x-y)^{\sigma}  \right] = 2 g^{\rho \sigma} \\
& \left[ (x-y)^{\rho}, (x-y)^{\sigma} \right] = 0\,.
\end{eqnarray*}
It is easily found that the subalgebra of functions
\begin{eqnarray*}
&M^{\rho \sigma}=\dfrac{(x+y)^{\rho}(x-y)^{\sigma}-(x-y)^{\rho}(x+y)^{\sigma}}{2}\\
&P^{\mu} = (x-y)^{\mu}
\end{eqnarray*} 
is invariant with respect ot the vector field $\tilde{\Gamma}$ and therefore is a Poisson subalgebra. Furthermore they obey the commutation relations of the Lie algebra of the Poincar\'{e} group and the corresponding Hamiltonian vector fields will provide us with a realization of this algebra as derivations of the associative product among functions. 

\section{A Lagrangian Jacobi structure and Peierls Bracket}
In this last section we will make a comparison between the covariant formalism presented in this letter by means of the Jacobi bracket, and the covariant formalism elaborated in \cite{Asorey}, where the principal ingredient is Peierls bracket. However, in order to make such a comparison, we have to move from Hamiltonian setting to the Lagrangian setting, i.e., from the cotangent bundle to the tangent bundle. Actually cotangent bundle has been used up to now because of the presence of the natural Liouville one form $\theta_0$ which is a potential for the canonical symplectic form $\omega = \sum_j dp_j\wedge dq^j$. The pull-back of $\theta_0$ to a particular codimension one submanifold of the cotangent bundle defines a contact structure. 

On the contrary the geometrical structure of tangent bundles does not allow for the definition of a one-form in a natural way and we are forced to introduce a Lagrangian function in order to derive the analog of $\theta_0$. In particular, coming back to the covariant description of relativistic particles, we can consider on $T\mathbb{R}^4$ the Lagrangian 
\begin{equation*}
\mathcal{L}=\sqrt{g_{\mu \nu}v^{\mu}v^{\nu}}\,, 
\end{equation*}
where $g_{\mu \nu}$ is the Minkowski metric tensor.
With this Lagrangian it is possible to associate the following one form \cite{Dubrovin,MFLMR}
$$
\theta_{\mathcal{L}} = \dfrac{\partial \mathcal{L}}{\partial x^{\mu}}dx^{\mu} = \dfrac{g_{\mu \nu} v^{\nu}}{\mathcal{L}} dx^{\mu}\,.
$$     

If we consider the submanifold $\Sigma$ such that $g_{\mu \nu}v^{\mu}v^{\nu} = 1$, the pull-back of $\theta_{\mathcal{L}}$ to $\Sigma$ defines a contact structure. The computations are analogous to the ones already presented in section 3: it is enough to replace $p_{\mu}$ with $v_{\mu}$ and fix $m^2 = 1$. We can also build up a two-point Jacobi bracket using the complete solution $S(x,y)$ presented in section 4.

On the other side the action functional 
\begin{equation}
S[\gamma] = \int_{\mathbb{R}}\mathcal{L} ds
\label{Rel Action}
\end{equation}  
permits to associate with this physical situation also another structure, Peierls bracket. Actually in the rest of this letter we will present a generalization of Peierls prescription \cite{Peierls} for the previous Lagrangian, which occurs to be singular. The main result is that this generalized Peierls bracket is part of a Jacobi structure defined on the space of functionals on the space of parametrized geodesics with tangent vectors of modulus one.  

Let us start from the equations of the motion. Euler-Lagrange equations associated with the action functional $\eqref{Rel Action}$ are 
\begin{equation}
\dfrac{d}{ds}\left( \dfrac{g_{\mu \nu} \dot{x}^{\nu}}{\mathcal{L}} \right) = 0
\end{equation}
and a solution is a unparametrized geodesic, which for our flat Minkowski space is given by an equivalence class of lines with constant velocities. For what is needed in the following part we can represent one solution by means of its momenta, which are 
$$
\dfrac{g_{\mu \nu} \dot{x}^{\nu}}{\mathcal{L}} = k_{\mu}\,.
$$ 

Following Peierls we select a reference solution $\gamma_0$ and linearize Euler-Lagrange equations around it. Considering the action functional $\eqref{Rel Action}$ we obtain the following Jacobi operator acting on a variation\footnote{As far as our analysis is concerned, we can consider variations of a given path $\gamma_0$ as described by vector fields along the path itself. However in a more general setting variation should be treated as homotopy classes of paths, a definition which can be extended also to situations where topology is non trivial. See for instance Ref. \cite{Zaccaria}} $\delta \gamma$
\begin{equation}
\mathcal{J}_{\gamma_0}\left[ \delta \gamma \right] = \dfrac{1}{\mathcal{L}}\left( g_{\mu \nu} - k_{\mu}k_{\nu} \right)\dfrac{d^2}{ds^2}\delta\gamma^{\nu} (s) = \dfrac{1}{\mathcal{L}}P_{\mu \nu} \dfrac{d^2}{ds^2}\delta\gamma^{\nu} (s).
\label{Lin Eq}
\end{equation}

Variations which are in the kernel of this operator (i.e. for variations which solve Jacobi equation) actually define the tangent space to any of the parametrized geodesic in the equivalence class of the solution $\gamma_0$. As the scalar product 
\begin{equation}
g_{\mu \nu} \delta \gamma ^{\mu} \dot{\gamma}^{\nu} = \left\langle \delta \gamma, \dot{\gamma} \right\rangle
\end{equation}
satisfies the equation
\begin{equation}
\dfrac{d^2}{ds^2}\left( \left\langle \delta \gamma, \dot{\gamma} \right\rangle \right) = 0 \,.
\end{equation}
It follows that $\left\langle \delta \gamma, \dot{\gamma} \right\rangle = as + b$ with $a,b$ two constants. Therefore a solution of Jacobi equation for the variations can be always decomposed into the orthogonal sum:
\begin{equation}
J(s) = J_{\perp}(s) + (as + b)\dot{\gamma}
\end{equation}
and $J_{\perp}(s)$ is such that $\left\langle J_{\perp} , \dot{\gamma} \right\rangle$ vanishes. Let us notice that this decomposition is possible only for time-like or space-like geodesics because in this case $\left\langle \dot{\gamma}, \dot{\gamma} \right\rangle = const \neq 0$. For light rays we have $\left\langle \dot{\gamma}, \dot{\gamma} \right\rangle=0$ and the component $J_{\perp}$ will have a non vanishing projection along $\dot{\gamma}$.

Let us fix now a given parametrization, for instance we will use proper time. This choice amounts to put $\left\langle \dot{\gamma} , \dot{\gamma} \right\rangle = 1$. Therefore solutions $J$ of Jacobi equation which are compatible with this ``gauge'' choice satisfy the following condition
$$
\left\langle \dfrac{d}{ds}J , \dot{\gamma} \right\rangle = \dfrac{d}{ds} \left\langle J, \dot{\gamma} \right\rangle = 0 \,,
$$
which implies that $\left\langle J, \dot{\gamma} \right\rangle$ is a conserved quantity along the geodesic. This allows to define the following one form on the space of geodesics:
\begin{equation}
\Theta(\gamma)[J] = \left\langle \dot{\gamma} , J \right\rangle
\end{equation}
which defines a contact structure on the space of parametrized geodesics with $\left\langle \dot{\gamma} , \dot{\gamma} \right\rangle = 1$ \cite{Bautista}. In particular the component $J_{\perp}(s)$ are in the kernel of this one form, or in other words they are the contact elements relative to this contact structure.
  
Let us now consider the following difference:
\begin{equation}
J^{\mu}_1 P_{\mu \nu}\dfrac{d^2}{ds^2} J_2^{\nu} - \dfrac{d^2}{ds^2}J^{\mu}_1 P_{\mu \nu}J_2^{\nu} = \dfrac{d}{ds} \left( J^{\mu}_1 P_{\mu \nu}\dfrac{d}{ds} J_2^{\nu} - \dfrac{d}{ds}J^{\mu}_1 P_{\mu \nu}J_2^{\nu} \right) \,.
\end{equation}
The left hand side of this equation vanishes when evaluated on $J$ which are solution of Jacobi equation, which implies the conservation of the expression in round brackets on the right hand side. This conservad quantity defines a two form $\Omega$ on the space of geodesics and we immediately see that this two form is degenerate, the kernel being made of variations $J(s) = (as+b) \dot{\gamma}(s)$. Therefore when we restrict this two form to a contact element we get a non degenerate antisymmetric bilinear operator. 

A simple argument allows to show that this two-form is actually the differential of the one form $\Theta$. Indeed if we fix a given value of the parameter, for instance $s=0$, we can associate with each geodesic its initial conditions $\left\lbrace \gamma^{\mu}(s=0) , \frac{d}{ds}\gamma^{\mu}(s=0) \right\rbrace = \left\lbrace x^{\mu}, v^{\mu} \right\rbrace$ and with a solution $J(s)$ of the Jacobi equation the initial conditions $\left\lbrace J^{\mu}(s=0) , \frac{d}{ds}J^{\mu}(s=0) \right\rbrace = \left\lbrace \frac{\partial}{\partial x^{\mu}}, \frac{\partial}{\partial v^{\mu}} \right\rbrace$. In these coordinates the one-form $\Theta$ is given by
$$
\Theta = g_{\mu \nu}v^{\mu} dx^{\nu}\,,
$$ 
whereas the two form $\Omega$ is written as
$$
\Omega = P_{\mu \nu} dv^{\mu} \wedge dx^{\nu} = (g_{\mu \nu} - v_{\mu}v_{\nu}) dv^{\mu} \wedge dx^{\nu}\,.
$$  
Since $v_{\mu}dv^{\mu} = 0$ on the submanifold $v_{\mu}v^{\mu}=1$, $\Omega$ coincide with the differential of $\Theta$. Actually from this expression we can immediately see the relationship with previous construction. The constraints that we have introduced in the previous sections can be seen as the constraints for the initial conditions of the solutions of a set of differential equations which come from a variational principle. Then the procedure outlined in this section permits to transport the geometrical structure of the initial conditon along the whole geodesic obtaining a description which does not require any splitting into space and time.

Up to now we have defined the contact form. In order to write the associated Jacobi bracket we will adapt Peierls' idea to our setting. Indeed let us consider a functional $A[\gamma]$ defined on the space of geodesics. One can associate with this functional a function of the finite dimensional manifold of the initial conditions. However we can avoid the splitting into space and time which is associated with the choice of a Cauchy surface and we can define a bivector field according to the procedure described in \cite{Asorey}. Therefore we can use the functional $A$ to set a new variational principle where the new action functional is given by $S+ \lambda A$. If we look for solutions which are perturbations of a given solution of the unmodified Euler-Lagrange equation, we get the Jacobi equation with a source term, that is
\begin{equation}
\dfrac{1}{\mathcal{L}}P_{\mu \nu} \dfrac{d^2}{ds^2}\delta_A\gamma^{\nu} (s) = - \dfrac{\delta A}{\delta \gamma ^{\mu}} \,.
\label{Per Jac Eq}
\end{equation}

Then we should select a solution $\delta_A\gamma^{\nu}(s)$ given in terms of the commutator Green function $\tilde{G}^{\mu \nu}(s-s')$ \cite{Sorkin} 
$$
\delta_A\gamma^{\nu}(s) = \int_{\mathbb{R}} ds' \tilde{G}^{\nu \mu}(s-s')\dfrac{\delta A}{\delta \gamma^{\mu}}(s')\,.
$$
Since the Jacobi operator has a kernel it is not possible to invert the linearized equation. However we can look for a right inverse of this operator, i.e. we can look for fundamental solutions satisfying the following equation
\begin{equation}
\dfrac{1}{\mathcal{L}} P_{\mu \nu}[k]\dfrac{d^2}{ds^2} G^{(\pm)\nu \rho}(s,s_0) = P^{\rho}_{\mu}[k]\delta (s,s_0)\,.
\end{equation} 

Eventually we get the following commutator Green function
\begin{equation}
\tilde{G}^{\mu \nu}(s,s') = \mathcal{L} P^{\mu \nu}[k](s-s')\,,
\end{equation} 
which is defined up to elements which are in the kernel of the Jacobi operator. When we restrict this operator to solutions that belong to a contact element it is no more degenerate and it coincides with the inverse of the two form $\Omega$, for any possible choice of the right inverse $\tilde{G}^{\mu \nu}(s,s')$. Therefore this is the bivector $\Lambda$ we need in the definition of a Jacobi bracket.

Let us now come back to the contact structure $\Theta$. The corresponding Reeb field is given by the solution of the Jacobi equation $J(s)= \dot{\gamma}$ as can be proven by a direct computation. Indeed we have that 
\begin{equation*}
\Theta(\gamma)[\dot{\gamma}] = \left\langle \dot{\gamma} , \dot{\gamma} \right\rangle = 1\,.
\end{equation*} 
Eventually we can write a Jacobi bracket between two functionals $A, \, B$ according to the following formula
\begin{equation*}
\left[ A\, , \, B \right](\gamma) = \int_{\mathbb{R}} \int_{\mathbb{R}} ds ds' \dfrac{\delta A}{\delta \gamma^{\mu}}(s)\tilde{G}^{\mu \nu}(s-s')\dfrac{\delta B}{\delta \gamma^{\nu}}(s') +
\end{equation*}
\begin{equation}
+ \int_{\mathbb{R}} \int_{\mathbb{R}} ds ds' A(s)\dot{\gamma}^{\mu}(s')\dfrac{\delta B}{\delta \gamma ^{\mu}}(s') - \int_{\mathbb{R}} \int_{\mathbb{R}} ds ds' B(s)\dot{\gamma}^{\mu}(s')\dfrac{\delta A}{\delta \gamma^{\mu}}(s')\,.  
\label{Jacobi Brackets for fields}
\end{equation}
This formula represents our generalization of the Peierls bracket in the form of a Jacobi bracket. The appearance of the vector field along the geodesics is reminescent of the eleventh generator considered by Dirac \cite{Dirac}.
A Poisson subalgebra is given by the set of functionals which are invariant under reparametrization of the geodesic, which are the functionals, say $A$, for which
$$
\int_{\mathbb{R}} ds \dot{\gamma}^{\mu}(s)\dfrac{\delta A}{\delta \gamma^{\mu}}(s) = 0\,.
$$ 

As an example we can compute the commutation relations between two functionals of the form
$$
A= \int_{\mathbb{R}} x^{\mu} \delta (s-s_1) ds \qquad B = \int_{\mathbb{R}} x^{\nu} \delta (s-s_2) ds
$$
and we get 
$$
\left[ A\, , \, B \right](\gamma) = P^{\mu \nu}(s_1 - s_2) + x^{\mu}(s_1)k^{\nu} - x^{\nu}(s_2)k^{\mu}
$$
and for $s_1 = s_2$ we get that 
$$
\left[ A\, , \, B \right](\gamma) = x^{\mu}k^{\nu} - x^{\nu}k^{\mu}
$$
which coincides with expression found in the previous sections, showing once more that space-time ``positions'' do not commute in this setting.

\section{Conclusions} 
In this work we have presented an alternative form for relativistic dynamics based on the use of Jacobi structures instead of Poisson structures on the algebra of observables. We have shown that it is possible to build a Jacobi bracket which describes the generalized dynamics (generalized in the sense of Dirac) for a particle of mass $m$ and which is manifestly invariant with respect to the action of Poincar\'{e} group. Indeed, according to this structure, the algebra of the generators of the Poincar\'{e} group is realized by means of Hamiltonian derivations of the associative product among observables and the corresponding generating functions close on a Poisson subalgebra. 

In the last section, which is a major contribution of this letter, we have proposed how to modify the usual Peierls prescription in order to treat also the case of the singular Lagrangian describing a test particle of mass $m$ in the flat Minkowski space-time. The final result has been the definition of a two-point Jacobi bracket on the space of time-like geodesics which does not require the introduction of a splitting into space and time. Actually this procedure can be generalized to non-flat spacetime and this will be dealt with in a forthcoming paper. 

The generalization we have proposed allows now for new questions which will be subject to further investigation in the near future. 

Firstly one could try to extend the procedure outlined for time-like particles to light-like particles. It is already known that the space of unparametrized light-like geodesics carries a contact structure. It could be interesting to understand if a modified version of the Jacobi bracket we have written could permit to reconstruct by suitable reductions \cite{Ibort} both the Poisson bracket on time-like unparamerized geodesics and to write a Jacobi bracket on the space of light rays. This could have deep consequences in the description of the propagation of electromagnetic fields in presence of gravitational fields.

A second question is related to description of gauge theories in a covariant way. According to a recent paper \cite{Balachandran} a generalization of Gauss law which would not require a splitting of spacetime into space and time should include the whole set of equations of motion. Then it would be interesting to extend our proposal in the last section to the case of gauge theories, first of all to electroodynamics. From the result of the last section we would not be surprised if an extension of the covariant description of gauge fields actually implied the definition of a Jacobi structure.

\section*{Acknowledgements}
This article is based upon work from COST Action  MP1405 QSPACE, supported by COST (European Cooperation in Science and Technology). This work has been also partially supported by the Spanish MICIN grant MTM2014-54692-P and QUITEMAD+, S2013/ICE-2801. M. Asorey  work has been partially supported by the Spanish MINECO/FEDER grant FPA2015-65745-P and DGA-FSE grant 2015-E24/2. G.Marmo would like to acknowledge the support provided by the Banco de Santander-UCIIIM ``Chairs of Excellence'' Programme 2016-2017.

\end{document}